\documentclass{article}
\usepackage{authblk}

\usepackage[table]{xcolor}

\usepackage[english]{babel}
\usepackage[utf8]{inputenc}
\usepackage{amssymb, amsthm, amsmath, amsfonts}

\usepackage{graphicx}

\usepackage{nicematrix, booktabs}
\usepackage[margin=1.5in]{geometry}

\usepackage[toc]{appendix}

\usepackage{float}

\usepackage[format=plain, labelfont=bf, textfont=it]{caption}

\newcommand*\rot{\rotatebox{90}}
\newcolumntype{s}{>{\columncolor[HTML]{E0E0E0}} c}

\begin{document}

\title{Survival stacking: casting survival analysis as a classification problem}

\author[1]{Erin Craig}
\author[2]{Chenyang Zhong}
\author[3]{Robert Tibshirani}
\affil[1]{Department of Biomedical Data Science, Stanford University}
\affil[2]{Department of Statistics, Stanford University}
\affil[3]{Departments of Biomedical Data Science and Statistics, Stanford University}

\date{\today}

\maketitle
\begin{abstract}
While there are many well-developed data science methods for classification and regression, there are relatively few methods for working with right-censored data. Here, we present ``survival stacking'': a method for casting survival analysis problems as classification problems, thereby allowing the use of general classification methods and software in a survival setting. Inspired by the Cox partial likelihood, survival stacking collects features and outcomes of survival data in a large data frame with a binary outcome. We show that survival stacking with logistic regression is approximately equivalent to the Cox proportional hazards model. We further recommend methods for evaluating model performance in the survival stacked setting, and we illustrate survival stacking on real and simulated data. By reframing survival problems as classification problems, we make it possible for data scientists to use well-known learning algorithms (including random forests, gradient boosting machines and neural networks) in a survival setting, and lower the barrier for flexible survival modeling.
\end{abstract}

\section{Introduction}

We consider the time-to-event setting of survival analysis, typically involving right censoring. To study when or whether an event occurs, we observe subjects over time, and we are rarely able to observe an entire co­hort until (1) they have the event of interest or (2) the study is complete. The standard survival model is the Cox proportional hazards model~\cite{cox1972regression}, which is not always appropriate: it is a linear model that assumes the relationship between the covariates and the hazard is constant through time. There is therefore a need for flexible survival analysis methods. Many popular methods -- boosting, random forests, and deep neural networks -- are well-developed for classification and regression, and less so for survival analysis. Further, there are few \emph{software packages} for survival analysis methods that support common properties of survival data, including time-dependent covariates and truncation.

In this paper we discuss  a method we call ``survival stacking''. This approach reshapes survival data -- including data with time-dependent covariates and truncation -- so that we can treat survival problems as classification problems, thereby enabling the use of classification methods in a survival setting. As a simple example, survival stacking converts the right-censored data set:

\[X = \begin{pNiceArray}{cc}[first-row]
\text{covariate $1$} & \text{covariate $2$}\\
x_{11} & x_{12} \\
x_{21} & x_{22}\\
x_{31} & x_{32}\\
\end{pNiceArray} 
\text{,}\hspace{.3cm} y = \begin{pNiceArray}{cc}[first-row]
\text{survival time} & \text{event}\\
t_1 & 1 \\
t_2 & 0 \\
t_3 & 1\\
\end{pNiceArray}
\]

to a ``survival stacked'' data set with a binary outcome:
\[ 
\widetilde{X} =\begin{pNiceArray}{cc | cc}[first-row]
\text{covariate $1$} & \text{covariate $2$} & \text{risk set $1$} & \text{risk set $2$}\\
x_{11}&x_{12}&1 & 0   \\
x_{21}&x_{22}&1 & 0  \\
x_{31}&x_{32}&1 & 0  \\
x_{31}&x_{32}& 0 & 1 \\
 \end{pNiceArray}
\text{,}\hspace{.3cm} \tilde{y} = \begin{pNiceArray}{c}[first-row] \text{outcome} \\ 1 \\ 0 \\ 0 \\1\end{pNiceArray}.
\]
This formulation is inspired by the Cox partial likelihood. And like the Cox partial likelihood, the survival stacking framework naturally supports time-varying covariates and truncation. We will motivate this framework intuitively, and we will describe in detail the relationship between survival stacking and the Cox model. 

By reframing survival problems as classification problems, we can now leverage the full suite of pre-existing software for classification and regression in a survival context, \emph{even in the presence of time-varying covariates and truncation}, and we are no longer restricted to the Cox proportional hazards assumption. As a result, survival stacking lowers the barrier for the development of innovative, flexible models for right censored data.

The idea of survival stacking is certainly not new, and  in Section~\ref{section:related} we discuss related work.
But we feel that it should be better known and more widely used.

The outline of this paper is as follows: Section~\ref{section:reshaping} gives a brief review of the Cox proportional hazards model, a description of the survival stacking method, and a theoretical analysis of the relationship between the two approaches. Section~\ref{section:general} describes the use of general classifiers with survival stacked data, and explains how to make predictions and perform model evaluation. Section~\ref{section:related} gives an overview of methods related to survival stacking, and Sections~\ref{section:realdata} and~\ref{section:simulated} give examples of survival stacking on real and simulated data.

\section{Survival stacking}
\label{section:reshaping}

\subsection{Review of the Cox model}
We consider survival data: in addition to covariates $x_i$, each subject $i$ has the outcome $(t_i, d_i)$, where $t_i$ is the last observation time for subject $i$, and $d_i$ indicates whether the subject experienced the event of interest at that time ($d_i=1$) or was lost to follow-up ($d_i=0$). We are interested in estimating the survival curve for a new subject, based on its covariates and the training dataset. Associated with each time $t$ is a set of subjects that are ``at risk'' at that time; these are the subjects who were not lost to follow-up before time $t$, nor did they have the event before time $t$. This is referred to as the \emph{risk set at time $t$}, noted here as $R(t)=\{\text{subject }j \mid t_j \geq t\}$. 

The standard method for survival analysis is the Cox proportional hazards model~\cite{cox1972regression}, which models the hazard, conditional on covariates $x$, as:
\begin{equation}
	\lambda(t \mid x) = \lambda_0(t) \exp\left(x^T \beta\right),
	\label{ref:coxhaz}
\end{equation}
where $\beta$ is a vector of coefficients, and $\lambda_0(t)$ is the baseline hazard that can be modeled flexibly. The coefficients $\beta$ are chosen through maximization of the \textit{partial likelihood}: 
\begin{align}
L_{\text{partial}}(\beta) &= \prod_{i : d_i = 1} P\left(\text{subject $i$ has the event}\mid \text{risk set $R(t_i)$}\right)\nonumber\\
&= \prod_{i : d_i = 1} \frac{\exp(x_i^T \beta)}{\sum_{j \in R(t_i)} \exp(x_{j}^T \beta)}\nonumber\\
\ell_{\text{partial}}(\beta) = \log(L_{\text{partial}}(\beta)) &= \sum_{i : d_i = 1}  \left[x_i^T \beta - \log\left(\sum_{j \in R(t_i)} \exp(x_{j}^T \beta)\right)\right].
\label{eqn:coxpl}
\end{align}

The partial likelihood is a product of conditional probabilities. At each observed event time ($t_i$ where $d_i = 1$), we include the probability that subject $i$ has the event, conditioned on the risk set at time $t_i$. In a sense, optimizing the partial likelihood is analogous to jointly optimizing a series of classification problems: at each event time, we wish to predict which member of the risk set had the event. This is the intuition underlying survival stacking.

\subsection{Survival stacking in detail}
Our goal is to reshape survival datasets to classification datasets so that we can treat survival problems as classification problems. As in the Cox partial likelihood, we consider a series of classification problems: at each observed event time, we construct a predictor matrix containing the covariates for each observation in the risk set at that time, and a categorical variable indicating the risk set. We also create a binary response vector indicating whether each member of the risk set had the event at that time. Again mirroring the Cox partial likelihood, we aim to jointly optimize these problems: we combine these data sets by vertically stacking them. 

We illustrate stacking with a small example dataset: 
\[X = \begin{pNiceArray}{cc}[first-row]
\text{covariate $1$} & \text{covariate $2$}\\
x_{11} & x_{12} \\
x_{21} & x_{22}\\
x_{31} & x_{32}\\
\end{pNiceArray} 
\text{,}\hspace{.3cm} y = \begin{pNiceArray}{cc}[first-row]
\text{survival time} & \text{event}\\
t_1 & 1 \\
t_2 & 0 \\
t_3 & 1\\
\end{pNiceArray}.
\]

Our dataset consists of three observations, each with two covariates. We observe a total of two events (the second observation is censored), and we define $t_1 < t_2 < t_3$. We begin by constructing a predictor matrix and binary response vector for the first observed event time, $t_1$. The risk set corresponding to $t_1$ is $\{1,2,3\}$, and so we use all three observations:
\begin{align*}
 \widetilde{X}_{R(t_1)} =\begin{pNiceArray}{cc|cc}[first-row]
\text{covariate $1$} & \text{covariate $2$} & \text{risk set $1$} & \text{risk set $2$}\\
x_{11}&x_{12}&1 & 0   \\
x_{21}&x_{22}&1 & 0  \\
x_{31}&x_{32}&1 & 0  \\
 \end{pNiceArray} & & \widetilde{y}_{R(t_1)}=\begin{pNiceArray}{c}[first-row] \text{outcome} \\ 1 \\ 0 \\ 0 \end{pNiceArray}.
 \end{align*}
 
 \noindent We repeat this for the second observed event time, $t_3$. The risk set at time $t_3$ is $\{3\}$, and so our predictor matrix and binary response vector are:
 \begin{align*}
 \widetilde{X}_{R(t_3)} =\begin{pNiceArray}{cc|cc}[first-row]
\text{covariate $1$} & \text{covariate $2$} & \text{risk set $1$} & \text{risk set $2$}\\
x_{31}&x_{32}& 0 & 1 \\
 \end{pNiceArray} & & \widetilde{y}_{R(t_3)}=\begin{pNiceArray}{c}[first-row] \text{outcome} \\ 1 \end{pNiceArray}.
 \end{align*}

\noindent Finally, we vertically stack our predictor matrices and response vectors to form a single dataset with a binary outcome:
\begin{align*}
 \widetilde{X} =\begin{pNiceArray}{cc|cc}[first-row]
\text{covariate $1$} & \text{covariate $2$} & \text{risk set $1$} & \text{risk set $2$}\\
x_{11}&x_{12}&1 & 0   \\
x_{21}&x_{22}&1 & 0  \\
x_{31}&x_{32}&1 & 0  \\
\specialrule{.005em}{0em}{0em} 
x_{31}&x_{32}& 0 & 1 \\
 \end{pNiceArray} & & \widetilde{y}=\begin{pNiceArray}{c}[first-row] \text{outcome} \\ 1 \\ 0 \\ 0  \\\specialrule{.005em}{0em}{0em} 
 1\end{pNiceArray}.
 \end{align*}

And we've finished reshaping our data! Our dataset now has a \emph{binary} outcome rather than a \emph{survival} outcome. When we model using the survival stacked data, we model the hazard at time $t$ conditioned on covariates $x$: the instantaneous rate of occurrence of the event.  

Survival stacking has a deep relationship to the Cox model: if we use logistic regression with our reshaped data, the coefficients will be a close approximation of those from the Cox model (see an example in Table~\ref{table:coefficients}). We describe this relationship more rigorously in Section~\ref{section:theory}.

\begin{table}[ht]
\centering
\begin{tabular}{rrr|rr}
\hline
& \multicolumn{2}{c|}{Coefficient} & \multicolumn{2}{c}{\textit{p}-value} \\
& Cox  & Logistic  & Cox  &  Logistic  \\
\hline
\text{age} & 0.01 & 0.01 & 0.11 & 0.11 \\
\text{grade} & 0.36 & 0.37 & 0.00 & 0.00 \\
\text{positive nodes} & 0.06 & 0.06 & 0.00 & 0.00 \\
\text{progesterone} & 0.00 & 0.00 & 0.31 & 0.31 \\
\text{estrogen} & 0.00 & 0.00 & 0.12 & 0.12 \\
\text{menopause} & 0.07 & 0.07 & 0.67 & 0.67 \\
\text{horm. treatment} & -0.28 & -0.28 & 0.02 & 0.02 \\
\hline
\end{tabular}
\caption{Example coefficients from the Cox proportional hazards model and logistic regression with the survival stacked data, using the Rotterdam Tumor Bank data set~\cite{royston2013external}, described in Section~\ref{section:realdata}.}
\label{table:coefficients}
\end{table}

Stacking naturally handles \emph{time-varying covariates} and \emph{truncation}. For time-varying covariates, use the appropriate covariates for each subject in each risk set when building the stacked matrix. To handle truncation, we simply include subjects in risk sets only when they have been observed in the data.

With survival stacking, we are not restricted to logistic regression: we can imagine using any two-class classifier! In a sense, survival stacking is a ``poor man's"  approach to the proportional hazards model --- it allows us to fit a model that accommodates censoring using simple software for binary classification. This opens up a new world of possibilities for survival modeling: we can now model the hazard using random forests, gradient boosting, and neural networks - and we can do so with familiar software for classification. We illustrate examples of survival stacking in Section~\ref{section:realdata}.

\subsection{Relationship between the Cox model and logistic regression with survival stacked data}
\label{section:theory}
There is an important relationship between the stacked binomial log-likelihood and the partial likelihood in the Cox model. The coefficients obtained by performing logistic regression on the stacked matrix are a close approximation of those obtained from the Cox proportional hazards model; this connection is also discussed by D'Agostino, Lee, et al. in \cite{d1990relation} (via a theoretical analysis, different than that included here), and by Ingram and Kleinman in \cite{ingram1989empirical} (through an example with real data). 

In the Cox proportional hazards model, when there is an event for subject $i$, the contribution of that event to the log partial likelihood is
\begin{equation}
\label{eqn:cox_theoretical}
    x_i^T \beta-\log\left(\sum_{j\in R(t_i)}\exp(x_j^T \beta)\right)
\end{equation}

\noindent Now, suppose that we treat the same event in a logistic regression model. Then, the contribution to the binomial log-likelihood is:
\begin{equation}
\label{eqn:lr_theoretical}
    \alpha_{t_i} + x_i^T \beta - \sum_{j \in R(t_i)}\log(1+\exp(\alpha_{t_i}+x_j^T \beta)),
\end{equation}
\noindent where $\alpha_{t_i}$ is the coefficient for the $t_i^\text{th}$ risk set indicator. Note that the logistic regression log-likelihood models the baseline hazard $\alpha_{t_i}$, while the partial likelihood does not. We optimize (\ref{eqn:lr_theoretical}) over $\alpha_{t_i}$ (by setting the partial derivative with respect to $\alpha_{t_i}$ to $0$)  to obtain:
\begin{equation*}
    \sum_{j \in R(t_i)}\frac{\exp(\alpha_{t_i}+x_j^T \beta)}{1+\exp(\alpha_{t_i}+x_i^T \beta)}=1.
\end{equation*}

\noindent If we use the approximation
\begin{equation}
1+\exp(\alpha_{t_i}+x_j^T \beta) \approx 1,
\label{eqn:approx}
\end{equation}

\noindent then we have
\begin{equation}
    \hat{\alpha}_{t_i}\approx -\log\left(\sum_{j \in R(t_i)}\exp(x_j^T \beta)\right).
    \label{eqn:baselinestart}
\end{equation}

\noindent Hence, the contribution to the binomial log-likelihood~(\ref{eqn:lr_theoretical}) is approximately
\begin{equation*}
    x_i^T\beta - \log\left(\sum_{j\in R(t_i)}\exp(x_j^T \beta)\right)-1,
\end{equation*}
which is the same as that for partial likelihood~(\ref{eqn:cox_theoretical}), up to a constant, and is identical to that for the profile likelihood (more detail is in Appendix~\ref{app:likelihoods}). The approximation (\ref{eqn:approx}) works best for the large risk sets and will err the most for events that occur near the end of the time period.

We note our approximation of $\hat{\alpha}_{t_i}$, the baseline hazard at $t_i$ (Equation~\ref{eqn:baselinestart}):
\[ 
\exp(\hat{\alpha}_{t_i}) \approx \frac{1}{\sum_{j \in R(t_i)}\exp(x_j^T \beta)}.
\]
This matches the Breslow estimate of the baseline hazard (Appendix~\ref{app:likelihoods}). Thus, when we do logistic regression using the stacked matrix, we jointly model the baseline hazard \emph{and} the coefficients~$\beta$ --- and the fitted values closely match those from the Cox regression and Breslow's estimate of the baseline hazard.
\\

Instead of logistic regression, we can use Poisson regression, where the form of the hazard is the same as that of the Cox model (assuming a discrete baseline hazard):
\[
\lambda(t \mid x) = \exp(\alpha_t + x^T \beta) = \exp(\alpha_t) \exp(x^T \beta).
\]
Likewise, the log-likelihood also matches the full Cox log-likelihood (Appendix~\ref{app:likelihoods}):
\begin{equation}
\sum_{i : d_i = 1}\Big[  \alpha_{t_i} + x_i^T \beta - \exp(\alpha_{t_i}) \sum_{j \in R(t_i)} \exp(x_j^T \beta) \Big].
\label{eqn:poissonll}
\end{equation}

\section{The use of general classifiers in the survival stacked setting}
\label{section:general}

\subsection{Choosing a learning algorithm, and handling time}
Having reshaped our data, we are ready to use general classification methods with the full survival stacked matrix. As we saw with logistic regression, linear models without interaction terms will preserve the proportional hazards assumption; coefficients for the risk set indicators act as the baseline hazard. Learning algorithms that discover interactions between features relax the proportional hazards assumption: risk set indicators may interact with the original data covariates, thereby allowing their influence on the hazard estimate to change across time. 

Different choices of learning algorithms have different virtues. For example, \texttt{glinternet}~\cite{lim2015learning} is a method that uses regularization to learn pairwise interactions in logistic regression. As a result, using \texttt{glinternet} with the stacked matrix is like doing Cox regression while also allowing the discovery of interactions between covariates, \emph{and} the discovery of time-varying effects (as interactions between the covariates and time). Non-linear methods naturally discover relationships between covariates, and between covariates and time: tree-based methods (random forests and gradient boosted trees) can discover interactions while neural networks can discover rich nonlinear relationships. 

To handle time, we may generalize the risk set indicators. Instead of treating time as a categorical variable, we may treat it as ordinal or continuous: when building the stacked matrix, we may include a single column containing the risk set \emph{time} in lieu of the risk set indicators. We may further generalize our representation of time using a basis expansion.

For large data sets, instantiating the full survival stacked matrix in memory may be unreasonable, and mini-batching may be required.

\subsection{Prediction and model evaluation}
We are often interested in predicting survival curves: at time $t$, the height of the survival curve gives the estimated probability of survival through time $t$. When we model using the survival stacked data, we estimate the hazard, $\hat{\lambda}(t \mid x)$. Equivalently, we estimate the conditional survival function, $1-\hat{\lambda}(t \mid x)$, the probability of surviving through time $t$, given survival up to time $t$. We can use the conditional survival function to estimate the survival curve:
\begin{equation}
\hat{S}(t \mid x) = \prod_{t_k \leq t} \Big(1 - \hat{\lambda} \left(t_k \mid x \right)\Big),
\end{equation}
as surviving through time $t$ requires surviving through all times before $t$.

The Cox proportional hazards model is often evaluated with Harrell's c-index~\cite{harrell1982evaluating}, which estimates whether the predicted hazard successfully ranks subjects in order of their true survival times. This is reasonable for models satisfying the proportional hazards assumption. However, for models that do not make the proportional hazards assumption, we must choose a different performance metric, as the model's ranking of subjects may change across time. To measure a model's discrimination, we recommend the time-dependent AUC~\cite{uno2007evaluating, gerds2013estimating}, and for calibration, we recommend the time-dependent Brier score~\cite{mogensen2012evaluating}, both computed using a meaningful time horizon. (If there is no single meaningful time horizon, we recommend the \emph{integrated} time-dependent AUC or Brier score.) To compute model performance metrics on a test data set, we first predict the survival curve at time $t$, $\hat{S}(t \mid x)$, for all test subjects. We then use the predicted $t$-year risk, $1-\hat{S}(t \mid x)$, in the computation of the AUC or Brier score.

\section{Related methods}
\label{section:related}
Here, we discuss methods related to survival stacking. In the context of logistic regression, the idea of survival stacking is not new. For example, Wu and Ware~\cite{wu1979use} present a model for the log-odds of the hazard: 
\[ \log\left(\frac{p(t \mid x(t))}{1-p(t \mid x(t))}\right) = a_t + g(x(t)), \]
where $x(t)$ describes the covariates (or a function of the covariates) at time $t$. When ${g(x(t)) = x(t)^T \beta}$, this matches the model obtained from logistic regression using the stacked matrix (allowing time-varying covariates). This approach is often referred to as pooled logistic regression, as employed and described in detail by Cupples, D'Agostino et al.~\cite{cupples1988comparison}. And the relationship between Cox regression and pooled logistic regression is well-studied: D'Agostino, Lee et al.~\cite{d1990relation} present a proof of the approximate equality between the Cox partial likelihood and the pooled logistic regression likelihood. (We presented a different proof here in Section~\ref{section:theory}.) Finally, as they describe the relationship between discrete and continuous proportional hazards models, Therneau and Grambsch~\cite{grambsch2000modeling} present a nice discussion of logistic regression with survival stacked data. After the first version of this article was written, a reader (Justin Max) pointed us to the excellent article from Allison~\cite{allison1982discrete}, which discusses the discrete approach in some detail.

\subsection{Discrete-time survival for neural networks}
The use of neural networks has been proposed for survival analysis: the most well-known among these include Cox-nnet~\cite{ching2018cox}, DeepSurv~\cite{katzman2018deepsurv} and RNN-surv~\cite{giunchiglia2018rnn}.
 
Here, we highlight Nnet-survival~\cite{gensheimer2019scalable}, due to its relationship with survival stacking. Nnet-survival first discretizes time, and then models the discrete-time hazard. Assuming no tied times, the contribution of time bin $[t_i, t_{i+1})$ (wherein a single event occurs) to the Nnet-survival loss function is: 
\[ \log(p_{[t_i, t_{i+1})}(x_i)) + \sum_{j: t_j \geq t_i\text{, }j \neq i} \log(1 - p_{[t_i, t_{i+1})}(x_j)),\]
where $p_{[t_i, t_{i+1})}(x_j)$ is the predicted hazard for individual $j$ during time $[t_i, t_{i+1})$. The hazard, $p_{t_i}$, comes from a neural network with a unique output node for each time point (and thus a unique bias, or baseline hazard, for each time point). As a result of this architecture, Nnet-survival naturally incorporates non-proportional hazards and a time-varying baseline hazard. 

The binning of time in Nnet-survival introduces a challenge: we must decide how to divide time into bins. Several rules of thumb are described in~\cite{gensheimer2019scalable}, but there is no definitive or obvious choice. Additionally, given a time bin $[t_i, t_{i+1})$, we must decide how to handle subjects censored during that window. Including those subjects in the loss term for this time bin may \emph{overestimate} their survival; likewise, excluding them \emph{underestimates} their survival. Gensheimer et al. recommend including censored subjects in all time bins where they were uncensored for at least half of the bin. Though they are nearly identical in spirit, time binning is one of the big differences between Nnet-survival and survival stacking: rather than discretize time into bins by choosing cut-points, survival stacking includes a term in the loss function for each observed event. 

\subsection{Multi-task learning in the context of survival analysis}
In the context of discrete survival analysis, we may consider the survival (or hazard) prediction at each time $t$ as a separate task; as in \emph{multi-task learning}~\cite{caruana1997multitask}, we may leverage knowledge across tasks.  

One example of multi-task learning in survival analysis is multi-task logistic regression (MTLR)~\cite{yu2011learning}. MTLR fits a generalization of the logistic regression model to predict a \emph{survival outcome} for each subject across discrete timepoints. By treating prediction at each time point as its own task, MTLR naturally models \emph{time-varying effects}. At each time $t_j$, MTLR finds a unique coefficient vector $\beta_j$; an L2 penalty is used to encourage smoothness in the coefficients across time. Unlike general multi-task learning approaches, however, the different tasks in MTLR are dependent, which is necessary to satisfy the monotone condition of survival functions. MTLR has been generalized neural-MTLR (N-MTLR)~\cite{fotso2018deep}, which uses a neural network in place of the linear model.


\section{Example with real data}
\label{section:realdata}
We illustrate our approach using the Rotterdam Tumor Bank and the German Breast Cancer Study Group data sets~\cite{royston2013external}. Both data sets are derived from studies of recurrence-free survival following primary surgery for node-positive breast cancer. We train and validate models with the Rotterdam Tumor Bank data ($1546$ observations, $38\%$ censored, median survival $24$ months), and we report results with the German Breast Cancer Study Group ($686$ observations, $57\%$ censored, median survival $21$ months). These data have 7 covariates: patient age, tumor grade, number of positive lymph nodes, measurements of progesterone and estrogen receptors, and binary indicators of hormonal treatment and menopause.

We compare common survival methods and survival stacking in terms of their time-dependent AUC and time-dependent Brier score, estimated at the $75^\text{th}$ percentile of observed event times. Among common survival methods, we consider the Cox proportional hazards model, a random survival forest~\cite{ishwaran2008random}, a gradient boosting machine with a Cox-based loss~\cite{freund1999short, friedman2001greedy} and Nnet-survival~\cite{gensheimer2019scalable}. Among the survival stacked approaches, we compare common classification methods: gradient boosting machine,  random forest, and a feed-forward neural network. We find that using \texttt{glinternet} with the survival stacked data achieves the best overall performance, balancing a high AUC and a low Brier score, and our full results are in Figure~\ref{image:performance}. 
{
\begin{figure}[H]
\centering
\includegraphics[width=.9 \linewidth]{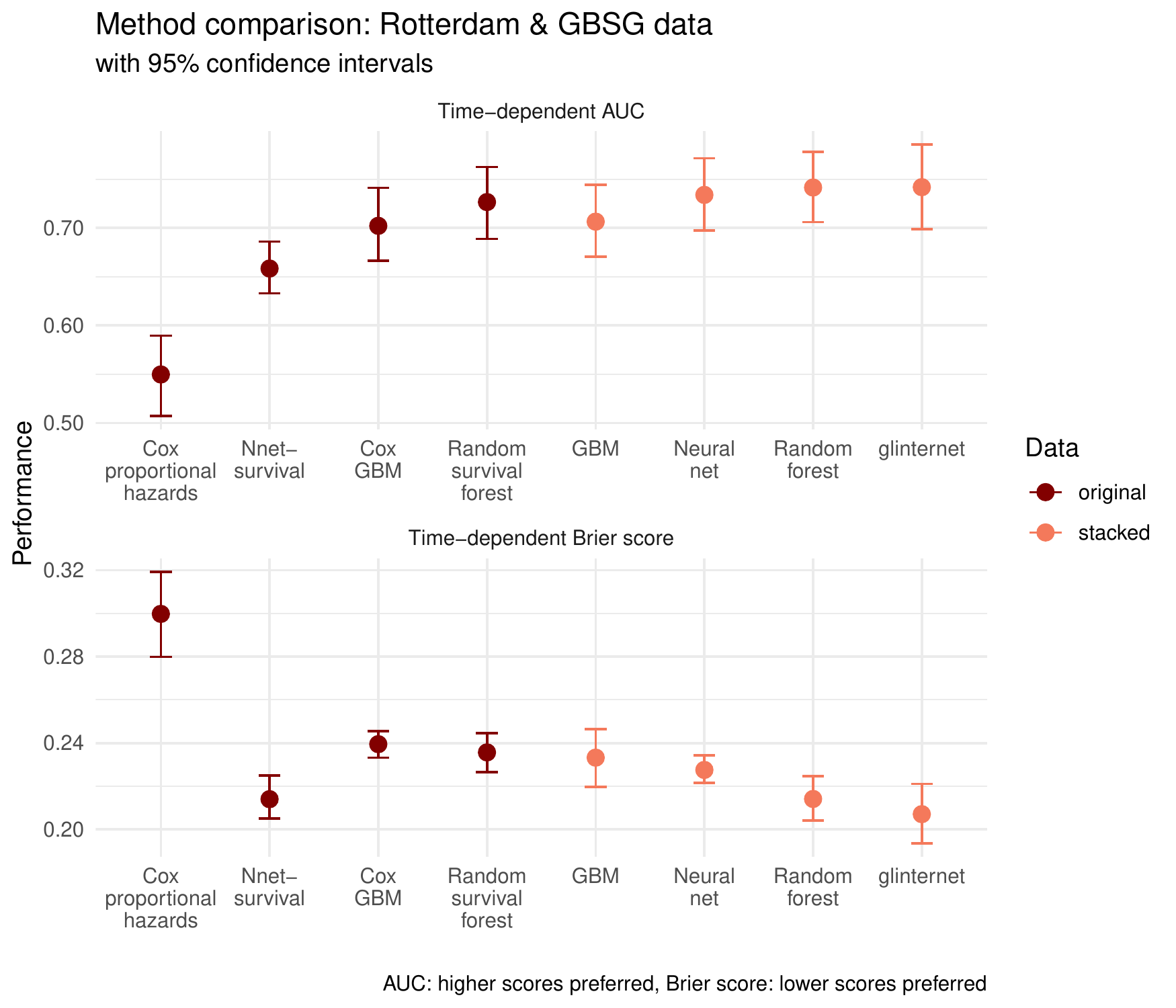}
\caption{Performance of models trained with the Rotterdam Tumor Bank data and tested with the German Breast Cancer Study Group data. We measure the time-dependent AUC and Brier score at the $75^\text{th}$ percentile of observed event times.}
\label{image:performance}
\end{figure}

\section{Performance on simulated data}
\label{section:simulated}
We simulate data where the hazard varies with time. We draw $N=3000$ subjects ($2000$ train, $1000$ test) with $p=5$ covariates from a standard normal distribution. To simulate a time-varying hazard, we discretize time into $10$ evenly spaced bins: $\{t_j\}_{j=1}^{10}$. The hazard at time $t_j$ is defined as \[h(t_j, x_i) = \frac{1}{1 + \exp(5 -\beta^T x_i - t_j \times \eta(t_j)^T x_i)}.\] We define $\beta = (-0.08,  -0.06, 0.02,  0, 0)$, and $\eta(t_j) = (5 \times (\frac{t_j}{10} - .25)^2-1, 0, 0, 0, 0)$ -- only the first covariate has a time-varying relationship with the hazard. Finally, at each time $t_j$, we determine whether the subject had the event by drawing from a binomial distribution (using the hazard as the event probability); the subject's event time is the first time this draw is $1$. Censoring times are drawn from an exponential distribution with rate $\frac{1}{5}$, and subjects who have not had the event by the final observation time are censored. We find that using \texttt{glinternet} with the survival stacked data achieves the best overall performance, though all of the survival stacked methods are competitive, and Nnet-survival and the random survival forest also perform well.
 
\begin{figure}[H]
\centering
\includegraphics[width=.9 \linewidth]{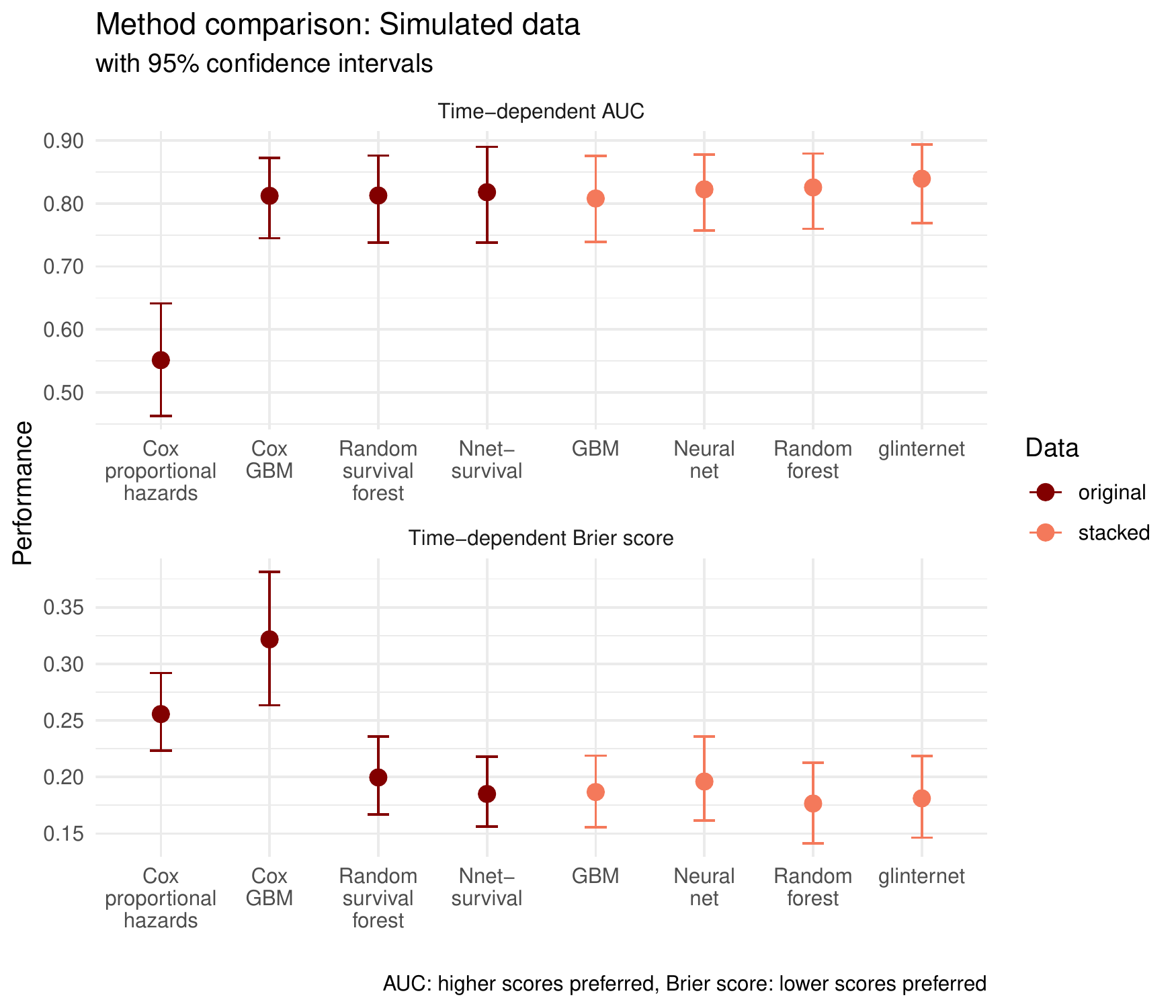}
\caption{Performance of models trained on simulated data, where the hazard is time-varying. We measure the time-dependent AUC and Brier score at the $75^\text{th}$ percentile of observed event times.}
\label{image:performance:simulated1}
\end{figure}

To illustrate the importance of correctly handling left-truncation, we repeat the above experiment, this time left-truncating half of the subjects in the training data. To left-truncate a subject, we define their ``start'' time using a draw from a random uniform distribution between $0$ and their event time. Again, survival stacking performs well: it flexibly supports the time-varying effect of the covariates on the hazard, and it appropriately handles the left-truncation.

\begin{figure}[H]
\centering
\includegraphics[width=.9 \linewidth]{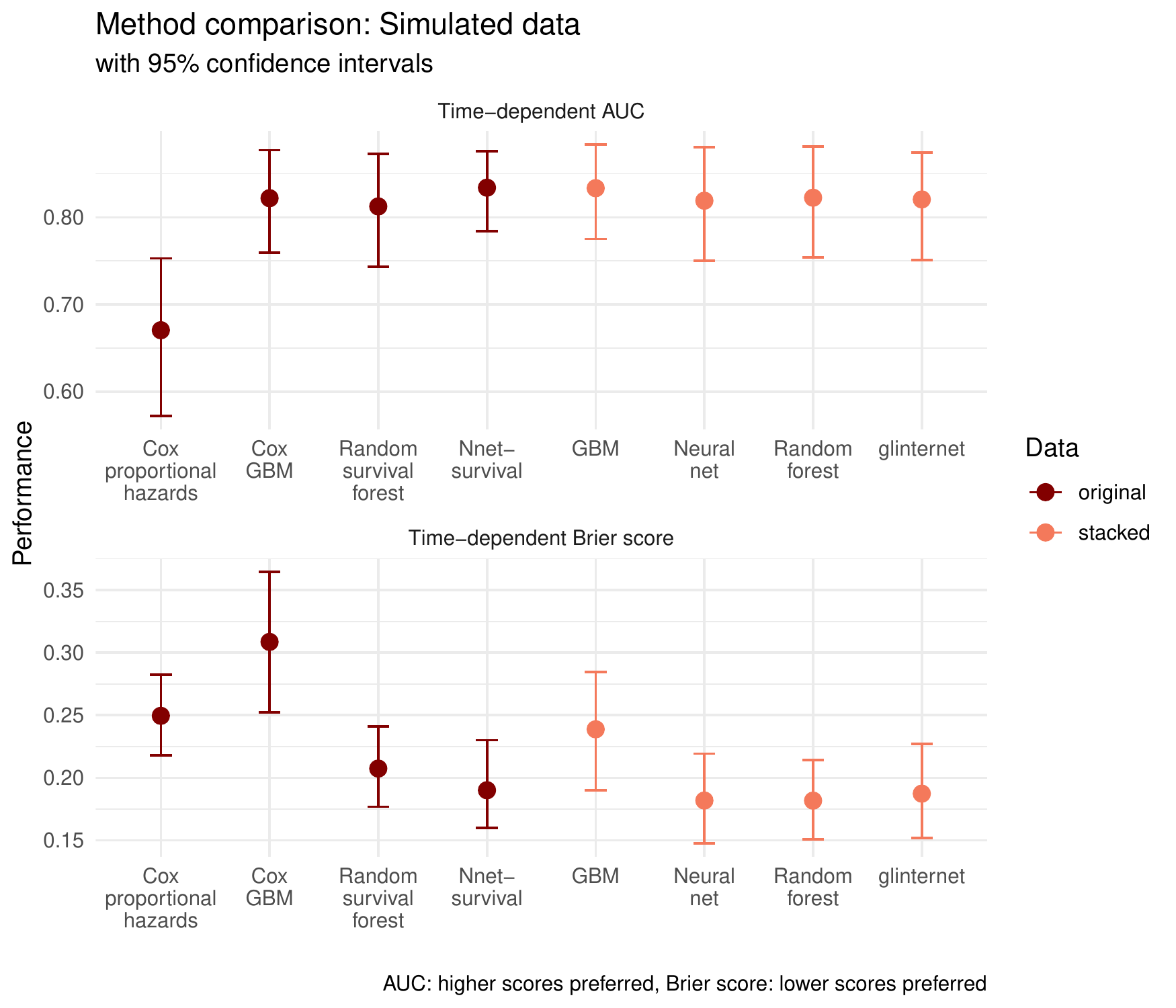}
\caption{Performance of models trained on simulated data, where the hazard is time-varying, and half of the training data is left-truncated. We measure the time-dependent AUC and Brier score at the $75^\text{th}$ percentile of observed event times.}
\label{image:performance:simulated2}
\end{figure}



\section{Conclusions and future directions}

Inspired by the Cox partial likelihood, survival stacking reframes survival problems as classification problems by reshaping survival data. Maximizing the Cox partial likelihood is analogous to jointly solving a series of classification problems: at each observed event time, we aim to predict which subject had the event. Survival stacking makes this explicit by constructing a classification problem for each event time, and combining them into a single data set, with an additional covariate indicating the event time. Unlike the Cox model, however, survival stacking does not require the proportional hazards assumption: the choice of classification method determines model flexibility. Further, survival stacking naturally handles time-varying covariates and truncation. 

Survival stacking may present challenges for larger data sets: for data with $n$ subjects, the survival stacked data has $O(n^2)$ rows. In the case of large data, mini-batching may be required. 

We will provide an R package with functions to (1) reshape data with a right-censored outcome to data with a binary outcome (to enable flexible survival modeling), and (2) transform model predictions to survival curves (to make it easier to evaluate model performance).   

Currently, the development of survival models is limited by the availability of flexible survival methods and software. This is particularly true for data with time-varying covariates, truncation or missingness, and for data that does not satisfy the proportional hazards assumption (as illustrated in Appendix~\ref{app:software}). Survival stacking makes flexible survival modeling accessible and straightforward by facilitating the con­ception and development of survival models using standard software for classification.

\section{Acknowledgements}
The authors thank Terry Therneau for the argument in Section~\ref{section:theory}, and we thank Terry Therneau, Thomas Gerds, Lu Tian, Trevor Hastie and Stephen Pfohl for helpful discussions. Robert Tibshirani was supported by NIH grant 5R01 EB001988-16 and NSF grant 19 DMS1208164.

\bibliographystyle{unsrt}
\bibliography{main}

%
%

\appendix

\section{The Cox partial and profile likelihoods}
\label{app:likelihoods}

Recall, the Cox partial likelihood is: 
\begin{align}
L_{\text{partial}}(\beta) &= \prod_{i : d_i = 1} P\left(\text{subject $i$ has the event}\mid \text{risk set $R(t_i)$}\right)\nonumber\\
&= \prod_{i : d_i = 1} \frac{\exp(x_i^T \beta)}{\sum_{j \in R(t_i)} \exp(x_{j}^T \beta)}\nonumber\\
\ell_{\text{partial}}(\beta) = \log(L_{\text{partial}}(\beta)) &= \sum_{i : d_i = 1}  x_i^T \beta - \log\left(\sum_{j \in R(t_i)} \exp(x_{j}^T \beta)\right).
\end{align}

For simplicity, we have assumed (and will continue to assume) that there are \emph{no tied times}: no two subjects have the event at the exact same time. Once fitted, the coefficients $\beta$ are often used to describe the \emph{relative risk} between subjects for different values of $x$. Optimizing the partial likelihood, however, does not allow us to say anything about the \emph{absolute risk} for any individual subject: the baseline hazard $\lambda_0(t)$ does not appear anywhere in the partial likelihood.  

To jointly model the baseline hazard, we can look at the full log-likelihood for the Cox model. We will assume that the baseline hazard is \emph{discrete}: the function $\lambda_0(t)$ takes values $\lambda_{t_1}, \lambda_{t_2}, \dots, \lambda_{t_k}$ at observed event times $t_1, \dots, t_k$, and $\lambda_0(t) = 0$ at all other times. The full log-likelihood for the Cox model is then:
\begin{equation}
\ell_{\text{full}}\big(\{\lambda_{t_i}\}_{i=1}^k, \beta\big) = \sum_{i : d_i = 1}\Big[  \log(\lambda_{t_i}) + x_i^T \beta - \lambda_{t_i} \sum_{j \in R(t_i)} \exp(x_j^T \beta) \Big],
\label{eqn:fullcoxll}
\end{equation}
where $t_i$ is the final observation time for subject $i$. 

We can use the full likelihood to estimate the baseline hazard as a function of $\beta$. We optimize the full likelihood (Equation~\ref{eqn:fullcoxll}) for $\lambda_{t_i}$ to obtain:
\begin{equation}
\lambda_{t_i}(\beta) = \frac{1}{\sum_{j \in R(t_i)} \exp(x_{j}^T \beta)}.
\label{eqn:breslow} 
\end{equation}

Equation~\ref{eqn:breslow} is known as Breslow's estimate of the baseline hazard~\cite{breslow1972discussion}, and it is the most common method of estimating the baseline hazard. We first estimate $\hat{\beta}$ by maximizing the partial likelihood, and then estimate the baseline hazard as $\{\lambda_{t_i}(\hat{\beta})\}_{i=1}^k$. 

Lastly, if we plug Equation~\ref{eqn:breslow} back in to the full likelihood, we obtain the \emph{profile likelihood}:
\begin{equation}
\ell_{\text{profile}}(\beta) = \sum_{i : d_i = 1} x_i^T \beta  -\log\Big(\sum_{j \in R(t_i)} \exp(x_{j}^T \beta)\Big) - 1,
\label{eqn:profilecoxll}
\end{equation}
which coincides with the \emph{partial likelihood} (Equation~\ref{eqn:coxpl}), up to a constant.

\section{Software: learning methods for survival data}
\label{app:software}

Survival stacking allows the modeling of survival data -- with time-varying covariates and truncation -- using linear and non-linear models, and it naturally enables the modeling of time-varying effects. Moreover, this is now possible using familiar, well-developed software for classification and regression. This is important, as there are very few \emph{survival software} packages that are equally flexible. 

Here, we examine the support for various features of survival data in common survival software packages in R and Python, with a focus on methods discussed in this work. We note that, though individual \emph{methods} may support a particular feature of survival data (e.g. Nnet-survival supports time-varying covariates), it is not always the case that the corresponding \emph{software} follows suit.

\begin{center}
\begin{table}[H]
	\begin{tabular}{ p{1.5in} p{1.4in} | s c s c s c }
		 \textbf{Package} & \textbf{Function} & \rot{ \textbf{Time dep.\ covs.} } & \rot{ \textbf{Truncation} } & \rot{ \textbf{Sample weights} } & \rot{ \textbf{Time varying effects} } & \rot{ \textbf{Non-linear} } & \rot{ \textbf{Missing data} } \\
		 \hline 
		 \rule{0pt}{4ex}\textbf{Linear models} & & & & & & \\
	 	 \texttt{survival}~\cite{survival-package} & \texttt{coxph} & $\checkmark$ & $\checkmark$ & $\checkmark$ & &\\ 
 		 \texttt{glmnet}~\cite{coxnet} & \texttt{glmnet} &  $\checkmark$ & $\checkmark$ & $\checkmark$ & &\\[2.5ex]
		 \hline  
		 
		 \rule{0pt}{4ex}\textbf{Random forests} & & & & & & \\
 		 \texttt{grf}~\cite{grf} & \texttt{grf} & & & & $\checkmark$ & $\checkmark$ & $\checkmark$ \\  
		 \texttt{ranger}~\cite{ranger} &  \texttt{ranger} & & & & $\checkmark$ & $\checkmark$ & $\checkmark$\\  
		 \texttt{randomForestSRC}~\cite{rfsrc} & \texttt{rfsrc} & & & $\checkmark$ & $\checkmark$ & $\checkmark$ & $\checkmark$\\  
		 \texttt{LTRCforests}~\cite{ltrc} & \texttt{ltrccif} & $\checkmark$ & $\checkmark$ &  & $\checkmark$ & $\checkmark$ & $\checkmark$\\
		  & \texttt{ltrcrrf} & $\checkmark$ & $\checkmark$ &  & $\checkmark$ & $\checkmark$ & $\checkmark$\\[2.5ex]  
		 \hline
		 
		 \rule{0pt}{4ex}\textbf{Boosting} & & & & & & \\ 
 		 \texttt{gbm}~\cite{gbm} & \texttt{gbm} & & &  $\checkmark$ & & $\checkmark$ & $\checkmark$ \\[2.5ex]  
		 \hline

		 \rule{0pt}{4ex}\textbf{Neural nets} & & & & & & \\
		 \texttt{survivalmodels}~\cite{survivalmodels} & \texttt{coxtime} & & & & $\checkmark$ & $\checkmark$ &\\  
		 &\texttt{deephit} & & & & $\checkmark$ & $\checkmark$ &\\  
		 &\texttt{deepsurv} & & & & $\checkmark$ & $\checkmark$ &\\  
		 &\texttt{loghaz} (Nnet-survival) & & & & $\checkmark$ & $\checkmark$ &\\  
		 &\texttt{pchazard} & & & & $\checkmark$ & $\checkmark$ &\\  
		 &\texttt{dnnsurv} & & & & $\checkmark$ & $\checkmark$ &\\[2.5ex]  
		 \hline  

	\end{tabular}
	\label{tab:Rsoftware}
\caption{Common \textbf{R} software packages, and their support for attributes of survival data.}
\end{table}
\end{center}

\begin{center}
\begin{table}[H]
\begin{tabular}{ p{1.5in} p{2.3in} | s c s c s c }
		 \textbf{Package} & \textbf{Function} & \rot{ \textbf{Time dep.\ covs.} } & \rot{ \textbf{Truncation} } & \rot{ \textbf{Sample weights} } & \rot{ \textbf{Time varying effects} } & \rot{ \textbf{Non-linear} } & \rot{ \textbf{Missing data} } \\
		 \hline 
		 \rule{0pt}{4ex}\textbf{Linear models} &&&&&& \\
	 	 \texttt{PySurvival}~\cite{pysurvival_cite} & \texttt{CoxPHModel} & & & & & &\\
		  & \texttt{LinearMultiTaskModel} & & & & $\checkmark$ & &\\[1ex]  
		  \texttt{lifelines}~\cite{davidson2019lifelines} & \texttt{CoxPHFitter} & & & & & &\\
		  & \texttt{CoxTimeVaryingFitter} & $\checkmark$ & $\checkmark$ & & & &\\[1ex]  
		   \texttt{scikit-survival}~\cite{sksurv} & \texttt{CoxPHSurvivalAnalysis} & & & & & &\\
		  & \texttt{CoxnetSurvivalAnalysis} & & & & & &\\[2.5ex]  
		 
		  \rule{0pt}{4ex}\textbf{Random forests} &&&&&& \\
		  \texttt{PySurvival} & \texttt{RandomSurvivalForestModel} & & & $\checkmark$ & $\checkmark$ & $\checkmark$ & $\checkmark$\\
		  & \texttt{ExtraSurvivalTreesModel} & & & $\checkmark$ & $\checkmark$ & $\checkmark$ & $\checkmark$\\
		  & \texttt{ConditionalSurvivalForestModel}  & & & $\checkmark$ & $\checkmark$ & $\checkmark$ & $\checkmark$\\[1ex] 
		  \texttt{scikit-survival} & \texttt{RandomSurvivalForest} & & & $\checkmark$ & $\checkmark$ & $\checkmark$ & $\checkmark$\\[2.5ex]  

		  \rule{0pt}{4ex}\textbf{Boosting} &&&&&& \\ 
		   \texttt{scikit-survival} & \texttt{GradientBoostingSurvivalAnalysis} & & & $\checkmark$ & $\checkmark$ & $\checkmark$ & $\checkmark$\\[2.5ex]  
		 
		  \rule{0pt}{4ex}\textbf{Neural nets} &&&&&& \\
		  \texttt{PySurvival} & \texttt{NeuralMultiTaskModel}  & & & & $\checkmark$ & $\checkmark$ &\\[1ex] 
		  \texttt{pycox}~\cite{kvamme2019time} & \texttt{CoxPH (DeepSurv)}  & & & & $\checkmark$ & $\checkmark$ & \\
		  & \texttt{LogisticHazard (Nnet-survival)}  & & & & $\checkmark$ & $\checkmark$ &\\
		  & \texttt{DeepHit}  & & & & $\checkmark$ & $\checkmark$ &\\
		  & \texttt{N-MTLR}  & & & & $\checkmark$ & $\checkmark$ &\\

		  \hline  
	\end{tabular}
	\label{tab:Pythonsoftware}
\caption{Common \textbf{Python} software packages, and their support for attributes of survival data.}
\end{table}
\end{center}

\end{document}